\documentclass[twocolumn,aps,prb,superscriptaddress,showpacs,floatfix]{revtex4-1}
\usepackage[utf8]{inputenc}
\usepackage{color}
\usepackage{amsmath}
\usepackage{amssymb}
\usepackage{graphicx}
\usepackage[unicode=true,pdfusetitle,
 bookmarks=true,bookmarksnumbered=false,bookmarksopen=false,
 breaklinks=false,pdfborder={0 0 1},backref=false,colorlinks=false]
 {hyperref}
\hypersetup{
 colorlinks,linkcolor=blue,citecolor=blue,urlcolor=blue}
\begin{document}

\title{Shubnikov-de Haas oscillations in the anomalous Hall conductivity
of Chern insulators}

\author{Luis M. Canonico}

\affiliation{Instituto de Física, Universidade Federal Fluminense, 24210-346 Niterói
RJ, Brazil}

\author{Jose H. García}

\affiliation{Catalan Institute of Nanoscience and Nanotechnology (ICN2), CSIC
and The Barcelona Institute of Science and Technology, Campus UAB,
Bellaterra, 08193 Barcelona, Spain}

\author{Tatiana G. Rappoport}

\affiliation{Instituto de Física, Universidade Federal do Rio de Janeiro, Caixa
Postal 68528, 21941-972 Rio de Janeiro RJ, Brazil}

\author{Aires Ferreira}

\affiliation{Department of Physics, York University,York YO10 5DD, U.K.}

\author{R. B. Muniz}

\affiliation{Instituto de Física, Universidade Federal Fluminense, 24210-346 Niterói
RJ, Brazil}
\begin{abstract}
The Haldane model on a honeycomb lattice is a paradigmatic example
of a system featuring quantized Hall conductivity in the absence of
an external magnetic field, that is, a quantum anomalous Hall effect.
Recent theoretical work predicted that the anomalous Hall conductivity
of massive Dirac fermions can display Shubnikov-de Haas (SdH) oscillations,
which could be observed in topological insulators and honeycomb layers
with strong spin--orbit coupling. Here, we investigate the electronic
transport properties of Chern insulators subject to high magnetic
fields by means of accurate spectral expansions of lattice Green's
functions. We find that the anomalous component of the Hall conductivity
displays visible SdH oscillations at low temperature. \textcolor{black}{The
effect is shown to result from the modulation of the next-nearest
neighbour flux accumulation due to the Haldane term,} which removes
the electron--hole symmetry from the Landau spectrum. To support
our numerical findings, we derive a long-wavelength description beyond
the linear ('Dirac cone') approximation. Finally, we discuss the dependence
of the energy spectra shift for reversed magnetic fields with the
topological gap and the lattice bandwidth. 
\end{abstract}

\maketitle

\section{Introduction}

Since its discovery the Hall effect has been the focus of keen interest
of researchers, particularly after the observation of its exactly
quantized version \cite{PhysRevLett.45.494}. \textcolor{black}{Thouless
}\textit{\textcolor{black}{et al}}\textcolor{black}{. \cite{PhysRevLett.49.405}
and Streda\cite{Streda82periodic} found that the noninteracting Hall
conductance }$\sigma_{xy}$\textcolor{black}{{} is a multiple of $e^{2}/h$,
as long as the Fermi energy lies inside a gap, even in Hall systems
with complex spectrum. T}hey derived a very interesting formula for
$\sigma_{xy}$, involving occupied Bloch states. Subsequently, it
was realized that the Hall conductance could be rewritten as $-e^{2}/h$
times a sum of Chern numbers associated with the filled bands \cite{PhysRevLett.51.2167},
which consist of the Berry curvatures \cite{Berry45} integrated over
the whole Brillouin zone. It then became clear that the Thouless-Kohmoto-Nightingale-den
Nijs (TKNN) formula for $\sigma_{xy}$ is a topological invariant,
and the integer Hall effect a robust topological property of the non-interacting
electron system. Some years later, Haldane \cite{PhysRevLett.61.2015}
proposed that the integer quantum Hall effect can occur in the absence
of Landau levels (LLs). He considered a single-orbital tight-binding
model on a honeycomb lattice \cite{PhysRevLett.53.2449} with a sublattice-staggered
on-site potential (orbital mass) and complex hoppings between next-nearest-neighbor
sites that produce a staggered magnetic field configuration with vanishing
total flux through the unit cell. The phase diagram of the model bornes
out two distinct topological phases surrounded by a conventional insulating
phase. Noninteracting systems hosting integer quantum Hall effect
in the absence of an external magnetic field are referred to as anomalous
quantum Hall insulators, or simply Chern insulators.

The advent of graphene \cite{Novoselov} and its remarkable properties
rekindled the interest in Haldane's predictions, encouraging both
the search for materials that would fulfill the key attributes of
his model, as well as inquires into alternative manifestations of
topologically protected states. Kane and Mele \cite{PhysRevLett.95.226801,PhysRevLett.95.146802,RevModPhys.82.3045},
for example, have shown that when spin-orbit interaction is taken
into account, it is possible to generate a quantum spin Hall phase
with conducting edge states that are protected against elastic backscattering
by time-reversal symmetry (TRS). The anomalous quantum Hall effect
was observed in thin films Bi(Sb)$_{2}$Te$_{3}$ doped with Cr \cite{Chang167},
and a few years ago the Haldane model was experimentally realized
using ultracold fermionic atoms in a periodically modulated optical
honeycomb lattice\cite{Jotzu2014}. Buckled honeycomb lattices (e.g.,
silicene) under in-plane magnetic fields are predicted to realize
the Haldane model requiring only the magnetic flux induced orbital
effect \cite{Wright13}. Furthermore, the possibility of an experimental
realization of Haldane's model have been invigorated by recent evidences
of strong proximity-induced SOC in graphene, \cite{Wang2015-WAL,Benitez2017,Ghiasi2017}
which together with the evidence of proximity-induced exchange interaction
in graphene on a ferromagnetic substrate \cite{PhysRevLett.114.016603,Hallal2017,Phong2017}
open realistic possibilities for future realizations of quantum anomalous
Hall effect in graphene.

Recently, Tsaran and Sharapov \cite{PhysRevB.93.075430} predicted
that two-dimensional systems of massive Dirac fermions exhibit strong
Shubnikov-de Haas (SdH) oscillations in the off-diagonal conductivity
that could be observed in the spin or valley Hall conductivity of
Dirac materials. Motivated by these studies, the present work employs
quantum transport simulations to explore the possible emergence of
SdHs in the anomalous (charge) Hall conductivity. For that purpose,
we use the kernel polynomial method (KPM) \cite{PhysRevE.56.4822,RevModPhys.78.275},
together with a numerical implementation developed by García \textit{et
al}. \cite{KPMtatiana-jose}, to calculate the off-diagonal conductivity.
Our numerical results for the Haldane model show visible quantum magneto-oscillations
in the anomalous component of $\sigma_{xy}$. However, differently
from Ref.~\cite{PhysRevB.93.075430}, the oscillations have origin
in an small asymmetry of high order Landau levels ($|n|>0$) under
field reversal $B\rightarrow-B$. Although such asymmetry seems to
have gone unnoticed in earlier works, we analytically show that it
derives from quadratic correction to the energy low-energy spectrum
around the Dirac points, and it can lead to sizable SdH oscillations
at low temperature. 

The article is organized as follows In section \ref{sec-si}, we review
the tight-binding Hamiltonian for the Haldane model in external magnetic
field, discuss the numerical method and report the SdH oscillations
in the anomalous part of the Hall conductivity. Section \ref{CL}
uses an extended low-energy approximation of the Haldane model to
analyze the dependence of the asymmetries with the model parameters.
In section \ref{conc} we summarize our findings and discuss how our
analysis can be used in transport experiments to characterize Chern
insulators.

\section{\label{sec-si}Model and Results}

We consider the Haldane model on a honeycomb lattice in a magnetic
field

\begin{align}
H=-\!\sum_{\langle i,j\rangle}t_{1}^{ij}c_{i}^{\dagger}c_{j}-\sum_{\langle\langle i,j\rangle\rangle}t_{2}^{ij}e^{\imath\varphi_{ij}}\,c_{i}^{\dagger}c_{j}+\sum_{i}M_{i}c_{i}^{\dagger}c_{i}\,,\label{eq:model}
\end{align}
where $t_{a}^{ij}=t_{a}e^{\phi_{ij}}$ are Peierls' substitution modified
hopping integrals with phases $\phi_{ij}=e/\hbar\int_{i}^{j}{\mathbf{A}\cdot d\mathbf{l}}$
, $\varphi_{ij}=\pm\varphi$ if an electron hops clockwise (anti-clockwise)
around a hexagonal plaquette\textcolor{black}{{} (Fig.\,1) }and $M_{i}$
are on-site potentials that equal $\pm M$ on sublattice $A$($B)$.
In what follows, we set $M=0$ and $\varphi=\pi/2$, so that the system
is a Chern insulator in the absence of external field ($B=0$). 
\begin{figure}[h]
\centering \includegraphics[width=0.96\columnwidth]{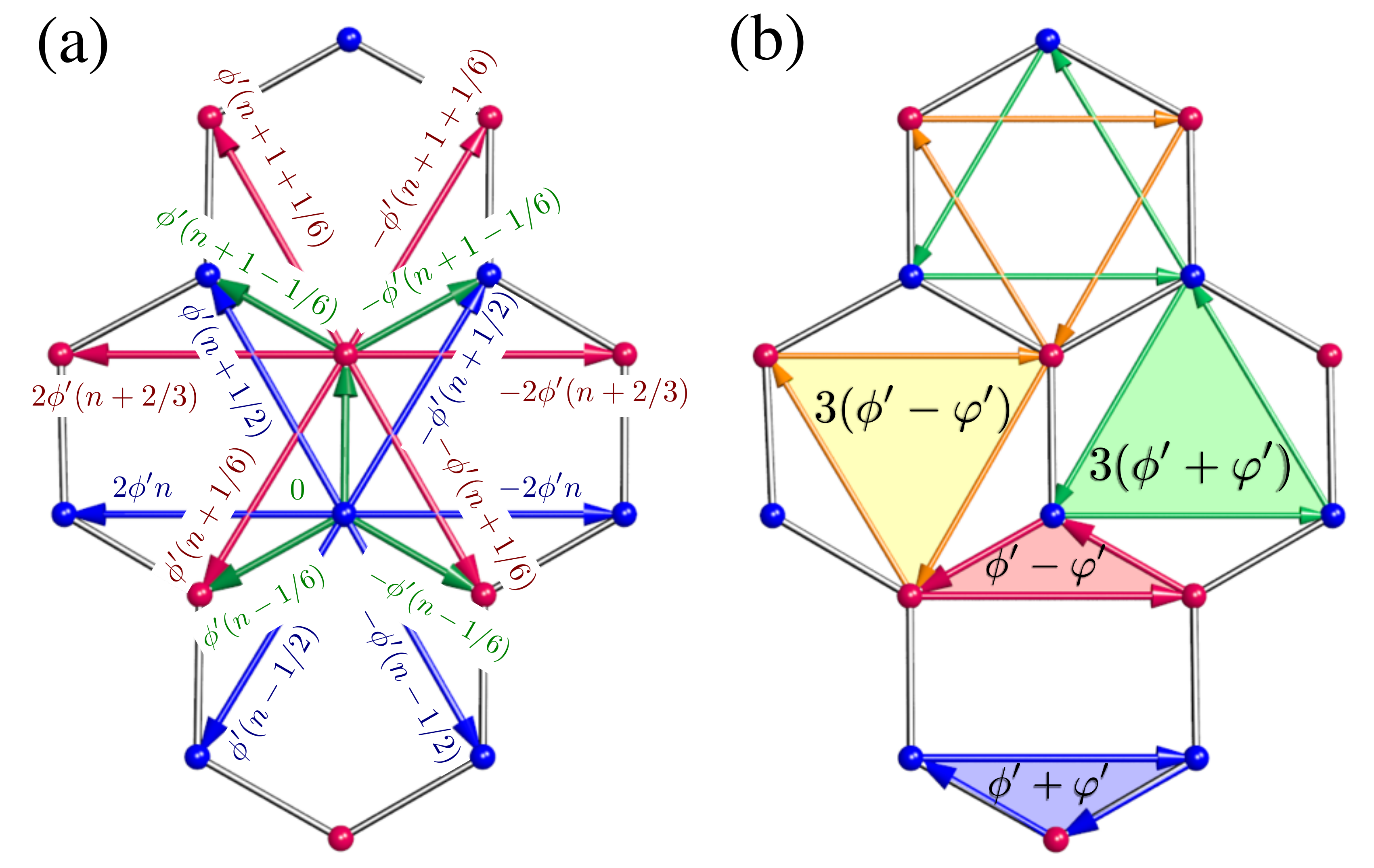} \caption{(a) Hopping phases of the Peierls substitution with a Landau gauge
$\mathbf{A}=-By\hat{x}$ in the unit cell$(n,m)$. Phases for the
NN hopping are in green. Phases for the NNN hopping in sublattices
A and B are in red and blue, respectively. Here, $\phi^{\prime}=3\sqrt{3}a^{2}eB/4\hbar.$
(b) Fluxes enclosed by the different paths involving NN and NNN hoppings
for positive $B$. }
\label{fig1} 
\end{figure}

In order to assess the density of states (DOS) and the transverse
conductivity of large systems we employ the KPM \cite{RevModPhys.78.275},
which has been extensively applied to investigate the electronic properties
of graphene layers \cite{Ferreira2011,Harju2014,Cysne2016,ReviewLeconte16}.
Within this approach the Green's functions and spectral operators
are approximated by accurate matrix polynomial expansions. Chebyshev
polynomials of first kind are the most popular choice given their
unique convergence properties and relation to the Fourier transform
\cite{boyd2001}. The expansion coefficients are computed by means
of a highly stable recursive procedure, which allows to treat very
large systems sizes. The first step is to rescale the energy spectrum
of Eq.\,(\ref{eq:model}) into the interval domain $[-1,1]$ of convergence
of the spectral series. This is easily achieved by defining rescaled
operators and energies variables, that is, $\tilde{H}=(H-b)/a$, and
$\tilde{E}=(E-b)/a$, where $a=(E_{T}-E_{B})/(2-\epsilon)$, and $b=(E_{T}+E_{B})/2$.
Here, $E_{T}$ and $E_{B}$ denote the top and bottom limits of the
energy spectrum, respectively, and $\epsilon$ is a small cut-off
parameter introduced to avoid numerical instabilities. To facilicate
numerical convergence, we follow Ref.\,\cite{KPMtatiana-jose} and
include Anderson disorder in $H$, with on-site energies $\epsilon_{i}$
randomly distributed in $[-\gamma/2,\gamma/2]$.

\begin{figure}[h]
\centering \includegraphics[width=0.45\textwidth]{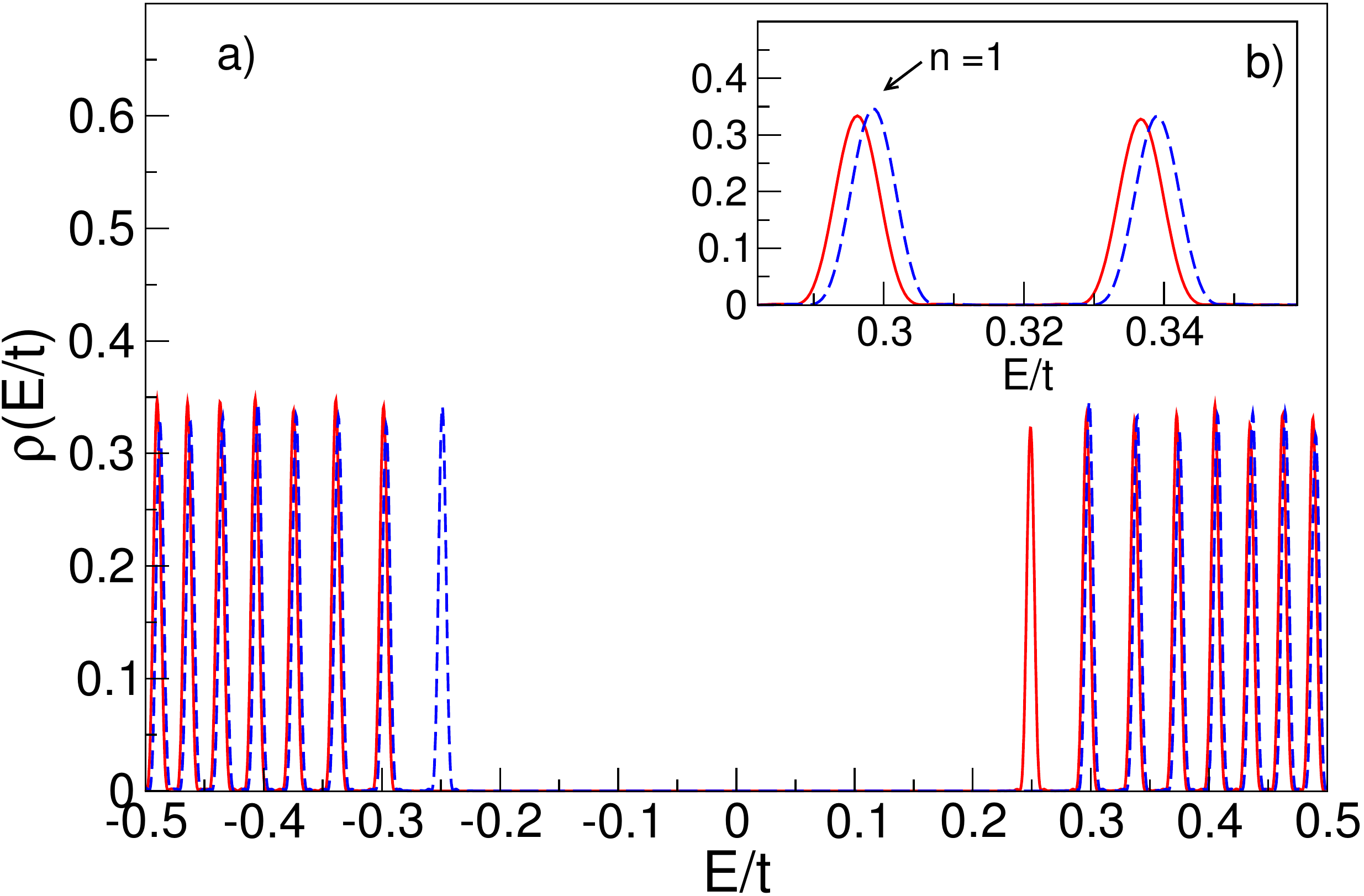} \caption{Density of states calculated as a function of energy for a system
described by the Haldane model with $2\times800\times400$ sites,
$t_{2}=\frac{0.5t}{6\sqrt{3}}$, $B=131T$ and $\gamma=0.1t$. The
spectral expansion {[}Eq.~(\ref{eq:DOS_expansion}){]} employed $4000$
polynomials and $70$ random vectors. A Jackson kernel was used to
damp Gibbs oscillations\cite{RevModPhys.78.275}. (a) Plot of the
full density of states for positive field configurations (blue dashed
line) and negative field configurations (red solid line). (b) Closeup
of the DOS around the $n=1$ and $n=2$ LLs.}
\label{fig:DOS} 
\end{figure}

The DOS of the system is expanded in terms of Chebyshev polynomials
$T_{m}(\tilde{E})=\cos(m\arccos(\tilde{E}))$. The $N$-order approximation
to the rescaled DOS is 
\begin{align}
\rho(\tilde{E})\simeq\frac{1}{\pi\sqrt{1-\tilde{E}^{2}}}\sum_{m=0}^{N-1}\mu_{m}g_{m}T_{m}(\tilde{E})\,,\label{eq:DOS_expansion}
\end{align}
where $g_{m}$ is a kernel introduced to damp spurious (Gibbs) oscillations.
The Chebyshev moments are obtained from $\mu_{m}=\textrm{Tr}\,\langle T_{m}(\tilde{H})\rangle$,
where $\langle...\rangle$ denotes disorder average. To reduce the
numerical complexity, we employ the stochastic trace evaluation technique
\begin{equation}
\mu_{m}\simeq\frac{1}{R}\left\langle \sum_{r=1}^{R}\langle\phi_{r}|T_{m}(\tilde{H})|\phi_{r}\rangle\right\rangle \,,\label{eq:mu_m}
\end{equation}
with complex random vectors $|\phi_{r}\rangle=D^{-1/2}\sum_{i=1}^{D}e^{\imath\theta_{i}}|i\rangle$,
where $\{|i\rangle\}_{i=1...D}$ is the original site basis and $\theta_{i}$
are independent random phases \cite{RevModPhys.78.275}. The DOS for
strong magnetic fields pointing along the $\pm\hat{z}$ directions
is shown in Fig.\,\ref{fig:DOS}. It reproduces the spectrum of the
Haldane model, as expected. The particle-hole symmetry breaking is
caused by the inclusion of the next-nearest neighbour hopping integral.
Panel (a) shows that the spectrum is only approximately symmetric
under reversal of the magnetic field direction. This is clearer in
panel (b) depicting a closeup of the DOS around the $n=1,2$ LLs,
where a small shift can be appreciated. Although the effect is relatively
small, it is not due to numerical inaccuracy. \textcolor{black}{The
assymmetry results from competing next-nearest neighbour flux accumulation
inside the plaquettes (see Fig.\,1) where closed loops connecting
sites of sublattice A have different phase variation than the ones
connecting sites of sublattice B that depends on the sign of the magnetic
field. This difference produces a mismatch between LLs of positive
and negative fields and it is responsible for the emergence of SdH
oscillations, as we shall subsequently see.}

To calculate the conductivity tensor $\sigma_{\alpha\beta}$, we use
an efficient numerical implementation of the KPM developed by García,
Covaci and Rappoport \cite{KPMtatiana-jose} based on the spectral
expansion of the Kubo-Bastin formula \cite{Bastin19711811}: 
\begin{align}
 & \sigma_{\alpha\beta}(\mu,T,B)=\frac{\imath e^{2}\hbar}{\Omega}\int_{-\infty}^{\infty}dE\,f(\mu,T,E)\nonumber \\
 & \times\textrm{Tr}\langle v_{\alpha}\delta(E-H)v_{\beta}\frac{dG^{+}}{dE}-v_{\alpha}\frac{dG^{-}}{dE}v_{\beta}\delta(E-H)\rangle.\label{eq:Kubo_Bastin}
\end{align}
In the above, $\mu$, $T$ and $B$ denote the chemical potential,
temperature, and applied magnetic field, respectively. The Cartesian
components of the velocity operator are designated by $v_{\alpha(\beta)}$,
with $\alpha,\beta=x,y$. $G^{\pm}$ stands for the retarded (advanced)
single-particle Green's function and $\delta(E-H)$ is the spectral
operator. Finally, $\Omega$ is the area and $f(\mu,T,E)=1/(1+\exp(-(\mu-E)/k_{B}T))$
is the Fermi-Dirac distribution function. The Green's functions and
spectral operators in Eq.\,(\ref{eq:Kubo_Bastin}) are expanded in
Chebyshev polynomials as performed for the DOS. Given the large number
of moments retained in our calculations, the energy resolution is
only limited by the mean level spacing $\delta E$ of the simulated
system \cite{ReviewLeconte16,CPGFmethod}. 
\begin{figure}[h]
\centering \includegraphics[width=0.45\textwidth]{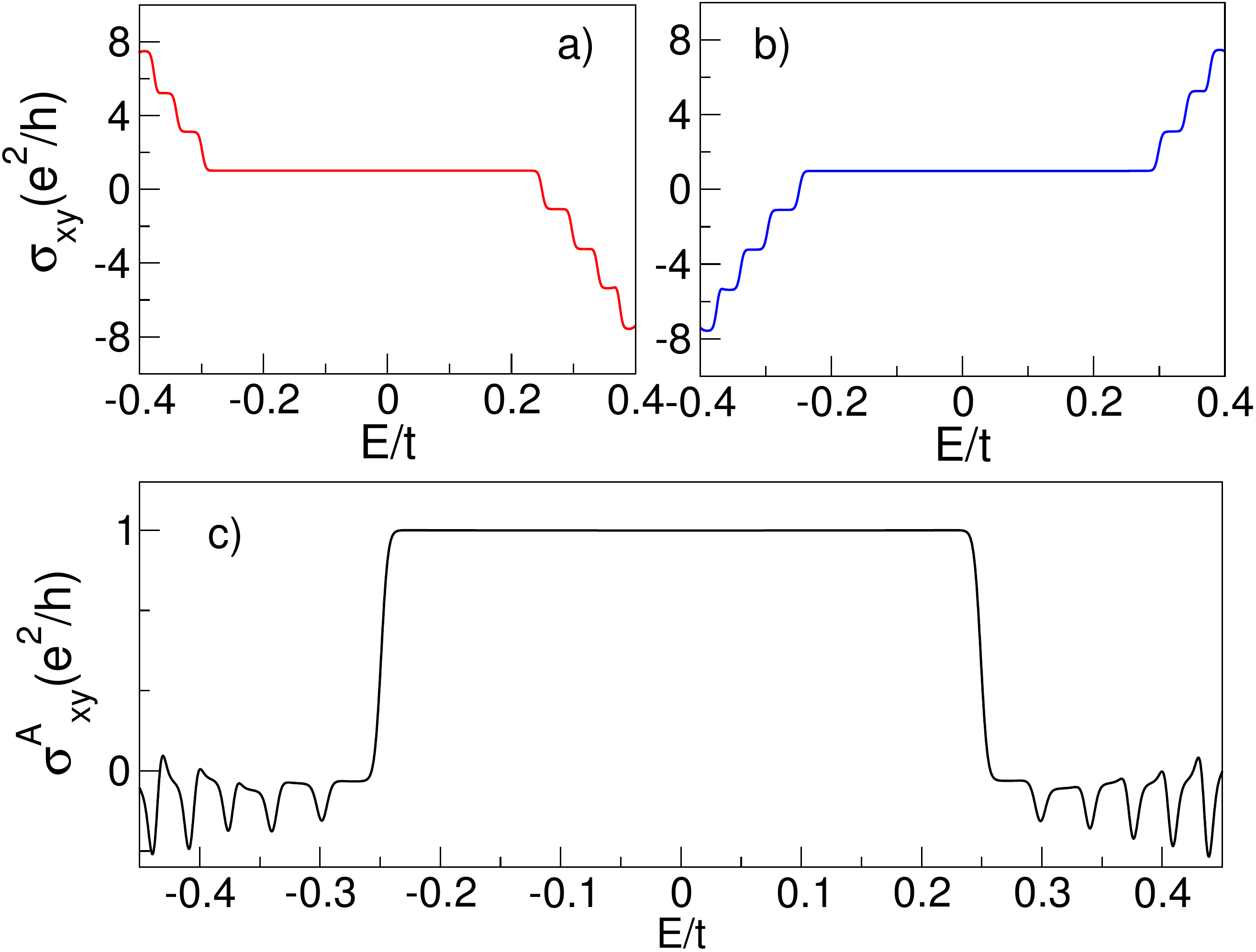} \caption{Hall conductivity of the Haldane model calculated as a function of
energy. Simulation parameters as in Fig.\ref{fig:DOS}. Panels (a)
and (b) depict the Hall conductivities calculated for $\vec{B}=\mp B\hat{z}$,
respectively, and panel (c) shows the calculated anomalous part of
the Hall conductivity. }
\label{fig:kpmconductivity} 
\end{figure}

The anomalous Hall conductivity consists ot two parts: (i) a regular
contribution $\sigma_{xy}^{R}$ anti-symmetric with respect to inversion
of magnetic field direction and (ii) an anomalous contribution $\sigma_{xy}^{A}$.
These are obtained from $\sigma_{xy}^{R(A)}=\frac{1}{2}\left(\sigma_{xy}^{+}\mp\sigma_{xy}^{-}\right)$
with $\sigma_{xy}^{\pm}=\sigma_{xy}(\mu,T,\pm B)$. The results of
our simulations are displayed in Fig.\,\ref{fig:kpmconductivity}.
The anomalous contribution $\sigma_{xy}^{A}$ to the Hall conductivity
is shown in panel (c). The steps in $\sigma_{xy}$ occur whenever
the energy crosses a LL (compare with DOS in Fig.~\ref{fig:DOS}).
The quantized anomalous Hall plateau is clearly visible, however oscillations
develop at high electronic density. Upon comparison with the DOS,
it becomes clear that the anomalous SdH oscillations are produced
by the shift in the spectra for $\vec{B}=\pm B\hat{z}$. These results
show that the breaking of electron--hole symmetry has important consequences
in the anomalous part of the Hall conductivity.\\

\section{\label{CL} Continuum Model}

To shed further light onto the anomalous oscillations seen in the
quantum transport calculations, we derive a low-energy continuum model.
To this end, we expand the tight-binding Hamiltonian Eq.\,(\ref{eq:model})
in momentum space around the inequivalent Dirac points $K_{\pm}$
in the Brillouin zone\cite{Ferreira2011}. The magnetic field is included
by minimal coupling. We choose the basis $(\mathbf{A},\mathbf{B})^{\textrm{t}}$
for the 4-component spinors with $\mathbf{A}=(A\,K_{+},A\,K_{-})$
(similarly for $\mathbf{B}$). To linear order in $\hbar\delta\mathbf{k}=\hbar(\mathbf{k}-\mathbf{K}_{\pm})$,
one obtains\cite{PhysRevLett.61.2015} 
\begin{align}
H_{\textrm{L.E.}}=v_{F}\left(\tau_{z}\otimes\pi_{x}\otimes\sigma_{x}+\pi_{y}\otimes\sigma_{y}\right)+\Delta\,\tau_{z}\otimes\sigma_{z},\label{eqn:continuum_zeroth}
\end{align}
The low-energy Hamoltonian describes the coupling between the momentum
of the particles and the pseudo-spin in the long-wavelenght limit.
$\vec{\pi}=\hbar\delta\vec{k}-e\vec{A}$ denotes the canonical momentum,
$v_{F}=3t_{1}a/2\hbar$ represents the Fermi velocity, $a$ is the
lattice constant, and $\tau_{z}=\pm1$ specifies choice of Dirac point,
$\mathbf{K}_{\pm}=\pm(4\pi/3a)\hat{k}_{x}$. Here, $\Delta=3\sqrt{3}t_{2}$
is referred to as the Haldane ``mass”. The spectrum reads as 
\begin{align}
{} & E_{n}^{(1)}=\eta\sqrt{\Delta^{2}+2|n|\left(\frac{\hbar v_{F}}{l_{B}}\right)^{2}},\hspace{0.5cm}\text{for \ensuremath{|n|\neq0\,,}} & {}\label{eq:En1}\\
{} & E_{0}^{(1)}=-s_{B}\Delta,\hspace{0.5cm}\text{for \ensuremath{n=0}.}\label{eq:E01}
\end{align}
$\eta=\pm1$ for electrons (holes), $s_{B}=\textrm{sign}(B)$ and
$l_{B}=(\hbar/e|B|)^{1/2}$ is the magnetic length. $E_{0}^{(1)}$
changes sign when the direction of the applied magnetic field is reversed.
However, for $|n|\neq0$, $E_{n}^{(1)}$ is independent of the field
direction, in contrast to the numerical results. This is true for
the expansion up to linear order in $\hbar\delta\vec{k}$, but the
inclusion of higher order terms can provided further refinements to
the LLs energy spectrum \cite{doi:10.1063/1.2981394,PhysRevB.88.165427}.
We then include quadratic terms in the low-energy expansion. As far
as the shift in the energy spectra for $\vec{B}=\pm\hat{z}$ is concerned,
it suffices to consider the correction to the next-nearest neighbours
hopping. We find 
\begin{align}
H_{\textrm{L.E.}}^{(2)}=H_{\textrm{L.E.}}-\frac{\boldsymbol{\pi}^{2}}{2m^{\prime}}\,\tau_{z}\otimes\sigma_{z}\,,\label{eqn:continuum_secod}
\end{align}
with $m'=2\hbar^{2}/(t_{2}9\sqrt{3}a^{2})$. The spectrum reads as
\begin{align}
{} & E_{n\neq0}^{(2)}=s_{B}\frac{\hbar^{2}}{2m'l_{B}^{2}}+\eta\sqrt{\left(\Delta-\frac{\hbar^{2}|n|}{m'l_{B}^{2}}\right)^{2}+2|n|\left(\frac{\hbar v_{F}}{l_{B}}\right)^{2}}\,, & {}\label{eq:En2}\\
{} & E_{0}^{(2)}=-s_{B}\left(\Delta-\frac{\hbar^{2}}{2m'{l_{B}}^{2}}\right),\hspace{0.2cm}\text{for \ensuremath{n=0}.}\label{eq:E20}
\end{align}
The inclusion of second order terms reproduces the LL shift when the
direction of magnetic field is inverted, as found in our numerical
simulations: $\Delta E_{n}^{(2)}=E_{n}(|B|)-E_{n}(-|B|)=\hbar^{2}/(2m'l_{B}^{2})$
for $|n|\ne0$ is independent of $n$, and increases linearly with
$t_{2}$. The inclusion of quadratic terms in the expansion also reduces
the effective contribution from the Haldane ``mass” by a factor that
increases linearly with $|n|$ for $|n|\ne0$. 
\begin{figure}[h]
\centering \includegraphics[width=0.4\textwidth]{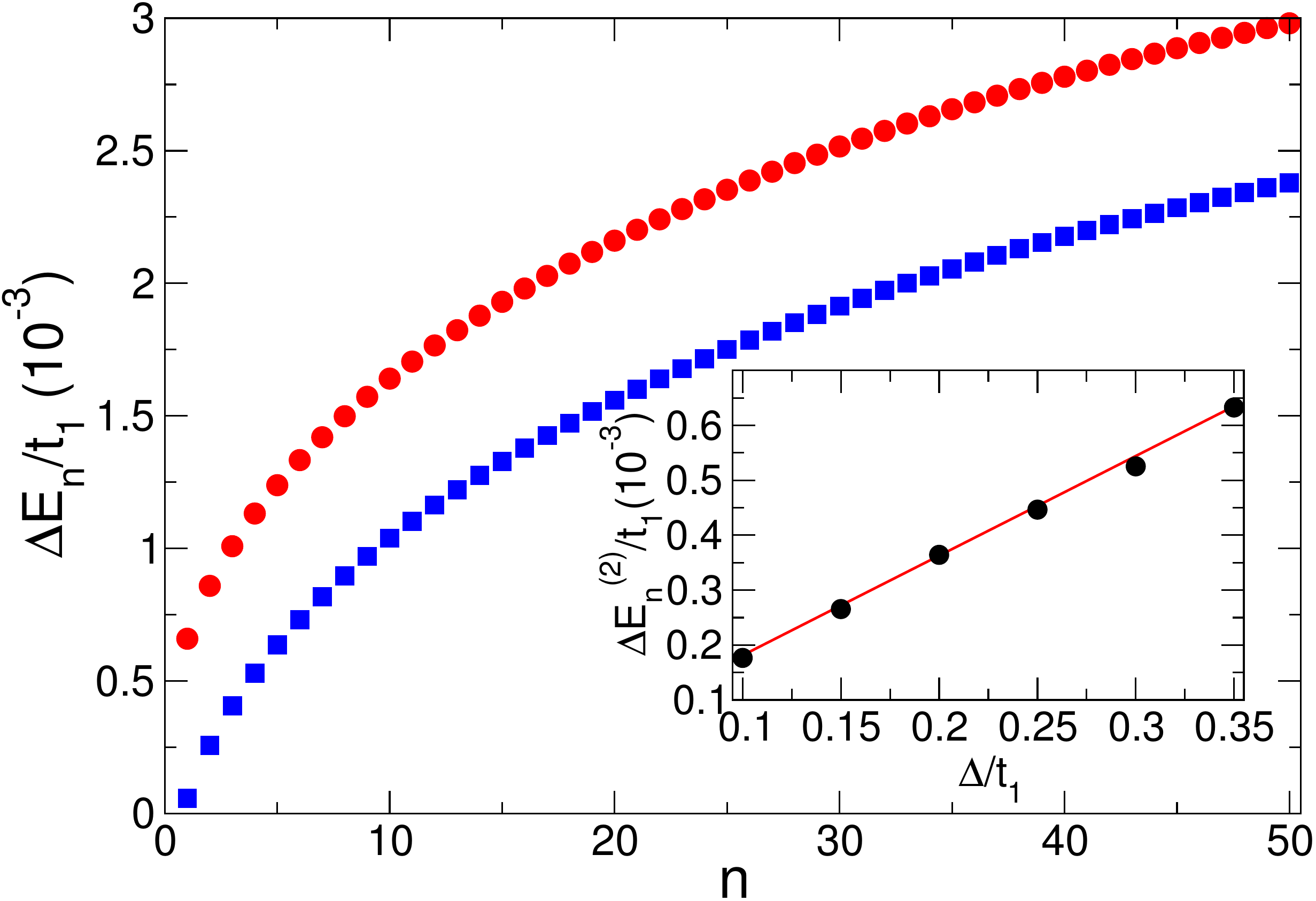} \caption{Energy spectra difference $\Delta E_{n}=E_{n}^{(1)}-E_{n}^{(2)}$
calculated for positive (open circle) and negative (solid circles)
magnetic field. Inset: Energy spectra shift according to continuum
model (blue dashed line) and tight-binding calculations (solid circles).
Simulation parameters: $t_{2}=\frac{0.2t}{6\sqrt{3}}$, $D=2\times10^{6}$
atoms, $|B|=157T$, $N=5000$ and $R=60$.}
\label{fig:LLspectra} 
\end{figure}

The difference between the energy spectra in the two approximations
is shown in Fig.~\ref{fig:LLspectra}. $\Delta E_{n}$ increases
monotonically with $n$ due to the character of the second-order correction
in Eq.\,(\ref{eqn:continuum_secod}). The dependence of the energy
shift with $n$ and $t_{2}$, which is related to the size of the
topological gap, can be used to extract the parameters of Chern insulators
in transport measurements.

In the inset of Fig.~\ref{fig:LLspectra} we compare the shifts determined
from Eq.~(\ref{eq:En2}) with the tight-binding results for different
values of the Haldane gap $\Delta=6\sqrt{3}t_{2}$. They are in excellent
agreement, showing that shift is a result of deviations from the linear
dispersion relation in the vicinity of the Dirac points, arising from
\textcolor{black}{competing}\textcolor{blue}{{} }next-nearest neighbour
hopping integral introduced by Haldane and the external magnetic field.\\

\section{Anomalous Oscillations }

The continuum model can provide crucial information on the anomalous
Hall conductivity for realistic magnetic fields not accessible with
our KPM implementation. For this purpose, we evaluate the transverse
conductivity $\sigma_{xy}$ within the empty-bubble approximation
\cite{PhysRevB.93.075430,PhysRevB.84.235410}

\begin{align}
\sigma_{xy}=\frac{e^{2}\hbar}{2\pi\imath\,l_{B}^{2}}\,\sum_{n\neq m}\,\frac{\langle v_{x}\rangle_{nm}\langle v_{y}\rangle_{mn}}{\delta E_{nm}^{(2)}\,(\delta E_{nm}^{(2)}+\imath\gamma)}\,\delta f_{nm}(\mu,T)\,,
\end{align}
where $E_{nm}^{(2)}=E_{n}^{(2)}-E_{m}^{(2)}$, $\delta f_{nm}$ is
the respective difference of occupation factors and $\gamma$ is a
broadening parameter. The velocity matrix elements for states around
$K_{+}$ are
\begin{align}
\langle v_{x}\rangle_{nm}=v_{F}N_{n,m}[\alpha_{n}\delta_{|n|,|m|-1}+\alpha_{m}\delta_{|n|-1,|m|}]\,,\label{eq:me1}\\
\langle v_{y}\rangle_{mn}=iv_{F}N_{n,m}[-\alpha_{n}\delta_{|m|-1,|n|}+\alpha_{m}\delta_{|m|,|n|-1}]\,,\label{eq:me2}
\end{align}
with
\begin{align}
 & \alpha_{n}=\frac{{E_{n}^{(2)}}-s_{B}\frac{\hbar^{2}}{2m'l_{B}^{2}}-\Delta+\frac{\hbar^{2}|n|}{{l_{B}}^{2}m'}}{\sqrt{2|n|}\frac{\hbar v_{F}}{l_{B}}}\,,\label{eq:b}\\
 & N_{nm}=\frac{1}{\sqrt{1+\left({\alpha_{n}}\right)^{2}}}\frac{1}{\sqrt{1+\left({\alpha_{m}}\right)^{2}}}.\label{eq:c}
\end{align}

\begin{figure}
\centering \includegraphics[width=0.9\columnwidth]{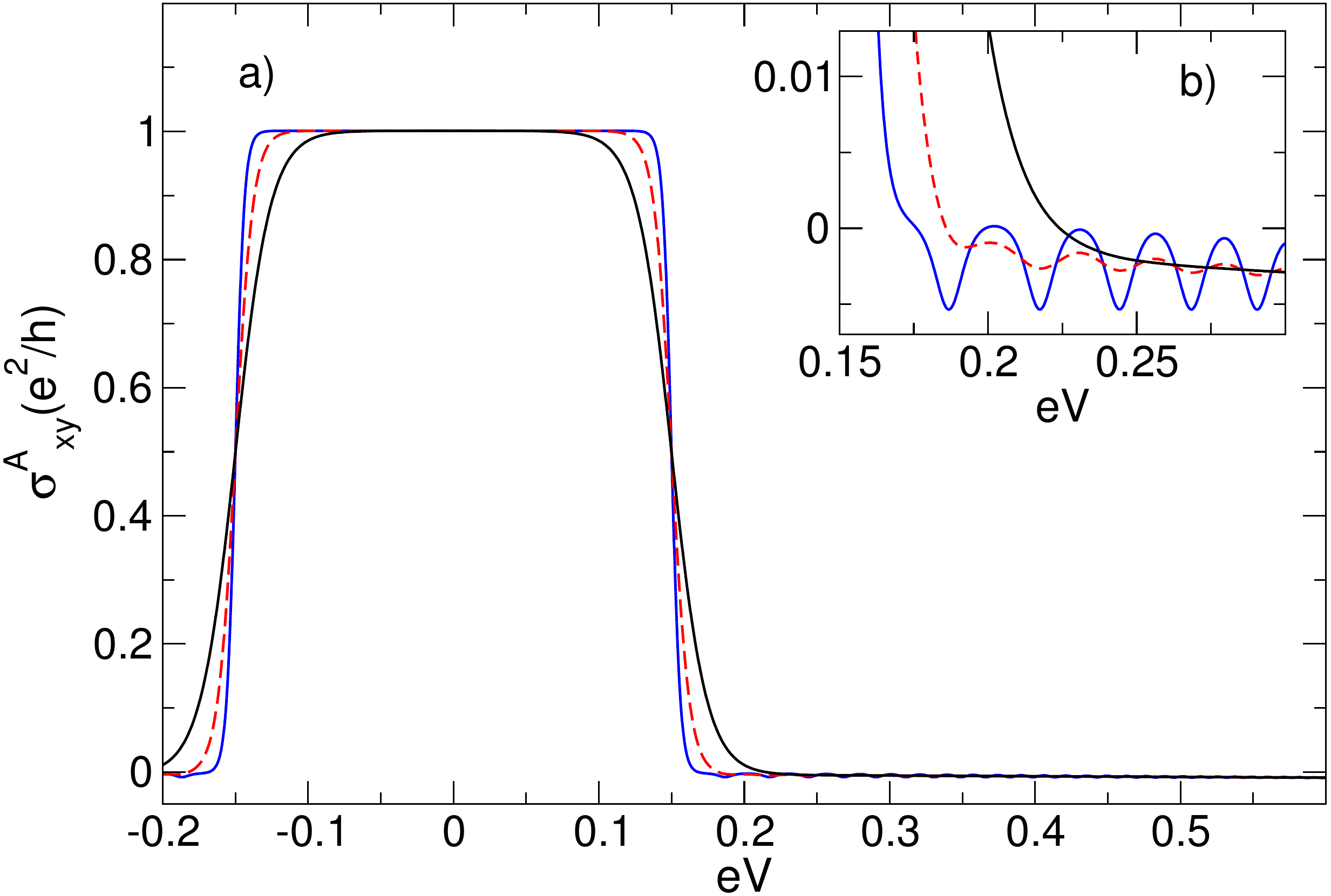} \caption{(a) Anomalous part of the Hall conductivity calculated for the Haldane
model in the continuum limit, including second order corrections.
Results are obtained for $t_{2}\approx0.03$ 
 eV, $|B|=10$ T, $\Gamma=0.3$ meV, and different temperatures: $T=35$
K (blue solid line), $T=70$ K (red dashed line), and $T=140$ K (black
solid line). 
Panel (b) highlights the SdH oscillations in the anomalous part of
the Hall conductivity}
\label{fig:cond_varioustreal} 
\end{figure}

Figure~\ref{fig:cond_varioustreal} shows the predicted anomalous
Hall conductivity at 10~T for selected temperatures, using indicative
values of hopping integrals motivated by a realization with graphene,
that is, $t_{1}=3$~eV and $t_{2}=3\times10^{-2}$~eV.\textcolor{black}{{}
For example, a Chern insulator could be induced in graphene by proximity
effect with either monolayer T'-WTe in combination with a ferromagnetic
insulator or magnetic doped T'-WTe. The WTe monolayer provides the
quantum spin Hall state~\cite{Wu76} and the spin degeneracy is lifted
by a ferromagnetic layer \cite{magnetic}. }The effect is less visible at higher temperatures
due to the smearing of the quantum Hall steps. \textcolor{black}{It
is noteworthy that these SdH oscillations have a different origin
from those discussed in Ref.~\onlinecite{PhysRevB.93.075430}, which
manifest in the valley or spin Hall conductivity and thus require
measurements of the nonlocal resistance\cite{Gorbachev448}. We deal
with a different model system, which in the absence of a magnetic
field is a Chern insulator with Chern number $C=1$. Here, the renormalized
Haldane ``mass” has opposite signs at $K_{\pm}$ and leads to a Hall
conductivity $\sigma_{xy}\ne0$ for $B=0$. The predicted anomalous
SdH oscillations should manifest in charge transport measurements
of $\sigma_{xy}(\mu,T,\pm B)$. }

\section{\label{conc}Conclusions}

We have investigated the transport properties of the Haldane model
in presence of strong magnetic fields by means of real-space calculations
and low-energy continuum models. We identified in our numerical calculations
a displacement between the energy spectra for magnetic fields of opposite
directions, verifying that the Landau levels for $|n|\neq0$ are only
approximately symmetric with respect to inversion of the applied magnetic
field $B\rightarrow-B$. \textcolor{black}{The mismatch between the
LLs of positive and negative magnetic fields }leads to SdH oscillations
in the anomalous contribution to the Hall conductivity, which can
be observed even at liquid nitrogen temperatures for Chern insulators
with large topological gaps. The presence of the quantum magneto-oscillations
in the anomalous contribution to the Hall conductivity arises as a
direct consequence of competing\textcolor{black}{{} neighbour flux accumulation}
due to broken TRS. Therefore, we expect this phenomenon to be present
in systems that exhibit an anomalous Hall state, such as magnetic
topological insulators, where the TRS is broken by magnetic ordering
\cite{Chang167,chang2015high-precision,PhysRevB.96.140410,PhysRevLett.114.187201,doi:10.1063/1.4983684}.
Furthermore, these oscillations could be used as a tool to extract
properties of the underlying microscopic mechanism that creates the
energy gap in the system, such as the next-nearest neighbor amplitude
in the Haldane model\cite{PhysRevLett.61.2015} or the tunneling amplitude
between surface states of thin films mediated by spin--orbit coupling
\cite{AHEinToplogicalMagneticInsulators,zhang2010crossover,kou2015metal-to-insulator}.
The recent observation of quantum spin Hall effect in two-dimensional
WTe$_{2}$ at temperatures of up to 100K~\cite{WTe1} hints at a
possible route for the fabrication of magnetic topological insulators
with large topological gaps, where SdH oscillations in the anomalous
contribution to Hall conductivity as described in this work could
be observed.

We acknowledge the Brazilian agencies CAPES and CNPq for financial
support. T.G.R. and A.F. acknowledge support from the Newton Fund
and the Royal Society through the Newton Advanced Fellowship scheme
(Ref. NA150043). J.H.G. received funding from the European Unions
Horizon 2020 research and innovation programme under grant agreement
No 696656 (Graphene Flagship). ICN2 is supported by the Severo Ochoa
program from Spanish MINECO (Grant No. SEV-2013-0295) and funded by
the CERCA Programme / Generalitat de Catalunya.

\vskip 5pt  \bibliographystyle{plainnat}
\bibliography{bibliografia}
 \bibliographystyle{apsrev} 
\end{document}